\documentclass[%
 reprint,
 amsmath,amssymb,
 aps,
prb,
]{revtex4-1}

\usepackage{graphicx}% Include figure files
\usepackage{dcolumn}% Align table columns on decimal point
\usepackage{bm}% bold math
\usepackage{epstopdf}
\begin{document}

\preprint{APS/123-QED}

\title{Critical current density,  vortex dynamics, and phase diagram of FeSe single crystal}% Force line breaks with \\

\author{Yue Sun$^{1}$}
\email[]{sunyue.seu@gmail.com}
\author{Sunseng Pyon,$^1$ Tsuyoshi Tamegai$^1$}
\email[]{tamegai@ap.t.u-tokyo.ac.jp}
\author{Ryo Kobayashi,$^2$ Tatsuya Watashige,$^2$ Shigeru Kasahara,$^2$ Yuji Matsuda,$^2$ and Takasada Shibauchi,$^3$}

\affiliation{%
 $^{1}$Department of Applied Physics, The University of Tokyo,Tokyo 113-8656, Japan \\
$^2$Department of Physics, Kyoto University, Kyoto 606-8502, Japan\\
$^{3}$Department of Advanced Materials Science, The University of Tokyo, Chiba 277-8561, Japan}

\date{\today}% It is always \today, today,

\begin{abstract}
We present a comprehensive study of the vortex pinning and dynamics in a high-quality FeSe single crystal, which is free from doping introduced inhomogeneities and charged quasi-particle-scattering because of its innate superconductivity. Critical current density, $J_c$, is found to be almost isotropic, and reaches a value $\sim$ 3 $\times$ 10$^4$ A/cm$^2$ at 2 K (self-field) for both $H$ $\|$ $c$ and $ab$. The normalized magnetic relaxation rate $S$ (= $\mid$dln$M$/dln$t$$\mid$) shows a temperature insensitive plateau behavior in the intermediate temperature range with a relatively high creep rate ($S$ $\sim$ 0.02 under zero field), which is interpreted in the framework of the collective creep theory. A crossover from the elastic to plastic creep is observed, while the fish-tail effect is absent for both $H$ $\|$ $c$ and $ab$. Based on this observation, the origin of the fish-tail effect is also discussed. Combining the results of $J_c$ and $S$, vortex motion in FeSe single crystal is found to be dominated by sparse strong point-like pinning from nm-sized defects or imperfections. The weak collective pinning is also observed and proved in the form of large bundles. Besides, the vortex phase diagram of FeSe is also constructed and discussed.

\begin{description}
\item[PACS numbers]
\verb+74.25.Wx+, \verb+74.25.Uv+, \verb+74.25.Sv+, \verb+74.70.Xa+

\end{description}
\end{abstract}

\pacs{Valid PACS appear here}% PACS, the Physics and Astronomy
\maketitle
\section{introduction}
As two representative members of the high-temperature superconductors (HTS), cuprate and iron-based superconductors (IBSs) share some similarities like layered structure, very high upper critical fields, doping phase diagram, and unconventional paring mechanism. \cite{StewartIBSsreview,AswathySUST} Also because of the higher operating temperatures, vortex motion and fluctuations are quite strong in both systems, which cause some very interesting phenomena in vortex dynamics like giant-flux creep, thermally activated flux flow, and fish-tail effect.\cite{Blatterreview} The vortex motion in HTS is determined by pinnings of two different types; the strong pinning attributed to sparse nanometer-sized defects\cite{vanderBeekPRB2002} and the weak collective pinning by atomic-scaled defects.\cite{FeigelmanPRL,Blatterreview} Therefore study of the vortex motion is a way to probe disorders in superconductors, which are related to both mechanism and application research.

Until now, most detailed studies on the vortex dynamics were performed on cuprates\cite{Blatterreview,ThompsonPRB,Yeshurunreview} and IBSs "122" phase\cite{ProzorovPRBBaCoJc,HaberkornPRBBaCo,HaberkornPRBCaNa,SalemSuguiPRB,TaenPRBBaCoirra,ShenBingPRB,TamegaiSUST} since high-quality single crystals are readily available. Through extensive research activities on the vortex system in cuprate superconductors, the collective creep theory has successfully interpreted novel features of the large creep rate and  plateau region in the temperature dependence of normalized relaxation rate ($S$ = $\mid$dln$M$/dln$t$$\mid$), and similar features have also been verified in IBSs.\cite{MalozemoffPRB,Yeshurunreview,ProzorovPRBBaCoJc,HaberkornPRBBaCo,HaberkornPRBCaNa,TaenPRBBaCoirra} However, some of the above materials possess complicated crystal structures, and all of them need further doping to introduce superconductivity. The complicated structure and element doping will easily cause inhomogeneities and defects in the crystal structure. Furthermore, most dopant atoms are charged such as hole-doping oxygen vacancies in YBa$_2$Cu$_3$O$_{7-\delta}$ (YBCO), electron-doping Co atom in Ba(Fe$_{0.93}$Co$_{0.07}$)$_2$As$_2$. Those charged dopant atoms will act as scattering centers for quasiparticles, i.e. weak pinning sites.\cite{vanderBeekPRL2010} All the inhomogeneities, defects, and scattering centers will jointly act as the pinning sources, and make the behavior of vortices complex. Thus, the study of vortex physics in crystal with simple structure without doping will be ideal to solve the puzzle.

FeSe is such a good candidate. It has the simplest crystal structure, composed of only Fe-Se layers, and shows superconductivity without further doping.\cite{HsuFongChiFeSediscovery} Recently, FeSe stimulated much interests since it is possible to break the superconducting transition temperature record ($T_c$ $\sim$ 55 K) in IBSs. Although the initial $T_c$ in FeSe is below 10 K,\cite{HsuFongChiFeSediscovery} it can be easily increased to over 40 K by intercalating space layers.\cite{BurrardNatMat,SunLilingnature} Furthermore, the monolayer of FeSe grown on SrTiO$_3$ even shows a sign of superconductivity over 100 K. \cite{GeNatMatter} For applications, high quality Te-doped FeSe tapes with transport $J_c$ over 10$^6$ A/cm$^2$ under self-field and over 10$^5$ A/cm$^2$ under 30 T at 4.2 K were already fabricated.\cite{SiWeidongNatComm} On the other hand, its relatively low $T_c$ is also meaningful to understand the crossover between HTS and low temperature superconductors, and to examine the validity of vortex-related theory in a much broader temperature range.

Unfortunately, the research of the vortex physics in FeSe is still left blank because of the difficulty to grow high-quality single crystal with enough size.\cite{HuRongweiPRB,McQueenFeSePRB,SBZhang} Recently, high-quality and sizable single crystals of FeSe have been grown.\cite{BöhmerPRB,ChareevCrystEngComm} In this report, we present a comprehensive study of the critical current density, vortex pinning, and dynamics in the high-quality FeSe single crystal. The normalized magnetic relaxation rate $S$ shows a temperature insensitive plateau with a relatively high creep rate, which is interpreted in the framework of the collective creep theory. A crossover from the elastic to plastic creep is also observed. Combining the results of $J_c$ and $S$, vortex motion in FeSe single crystal is proved to be dominated by sparse strong point-like pinning from nm-sized defects. The weak collective pinning is also observed and proved in the form of large bundles. Besides, the vortex phase diagram of FeSe was also constructed and discussed.

\section{experiment}
High-quality single crystals of tetragonal $\beta$-FeSe were grown by the vapor transport method as described elsewhere.\cite{KasaharaPNAS} Energy dispersive X-ray spectroscopy (EDX) shows the composition ratio of Fe to Se is $\sim$ 0.995, which is consistent with the structural refinement.\cite{KasaharaPNAS} Our previous scanning tunneling microscope (STM) result also proved that the crystal contains extremely small amount of impurities and defects (less than one impurity per 2000 Fe atoms).\cite{KasaharaPNAS} Magnetization measurements were performed using a commercial SQUID magnetometer (MPMS-XL5, Quantum Design). Resistivities were measured on the crystal with size of 800 $\mu$m $\times$ 350 $\mu$m $\times$ 35 $\mu$m by the four-lead method with a Physical Property Measurement System (PPMS, Quantum Design). In order to decrease the contact resistance, we sputtered gold on the contact pads just after the cleavage, then gold wires were attached on the pads with silver paste, producing contacts with ultralow resistance ($<$100 $\mu\Omega$).

\section{results and discussion}
\begin{figure}\center
\includegraphics[width=8.5cm]{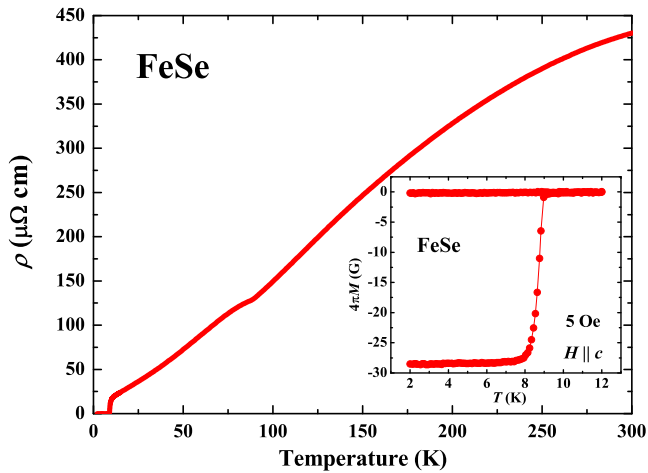}\\
\caption{(Color online) Temperature dependence of the resistivity from 300 - 2 K for FeSe single crystal. The inset shows the temperature dependences of ZFC and FC magnetizations at 5 Oe.}\label{}
\end{figure}

The inset of Fig. 1 shows the temperature dependence of zero-field-cooled (ZFC) and field-cooled (FC) magnetization at 5 Oe for FeSe single crystal. The crystal displays a superconducting transition temperature, $T_c$ $\sim$ 9 K defined as the onset of the separating temperature for FC and ZFC curves, which is higher than $T_c$ $\sim$ 8 K of early reports.\cite{BraithwaiteJPCM} Taking the criteria of 10 and 90\% of the magnetization result at 2 K, the superconducting transition width, $\Delta$$T_c$, is estimated as $\sim$ 0.6 K. The main panel of Fig. 1 shows the temperature dependence of in-plane resistivity for FeSe single crystal. An obvious kink behavior can be observed at a temperature about 90 K, which is already proved to be related to the structural transition at $T_s$. The residual resistivity ratios RRR, defined as $R$(300 K)/$R$(10 K), is close to 30, which is nearly one order larger than earlier samples.\cite{BraithwaiteJPCM} The higher $T_c$, small $\Delta$$T_c$, and large RRR all manifest the very high-quality of our single crystal. The results together with the extremely small amount of impurities proved by STM,\cite{KasaharaPNAS} demonstrate that our vortex physics study is performed in a clean crystal with less influence from impurities or inhomogeneities.

\begin{figure}\center
\includegraphics[width=8.5cm]{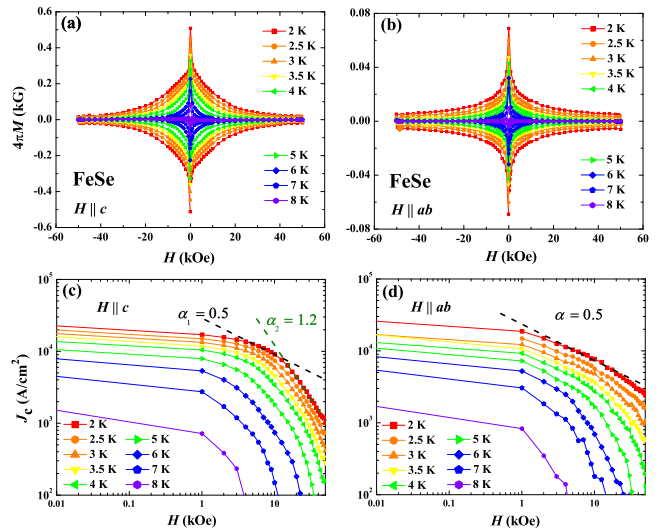}\\
\caption{(Color online) Magnetic hysteresis loops of FeSe single crystal at different temperatures ranging from 2 to 8 K for (a) $H$ $\|$ $c$ and (b) $H$ $\|$ $ab$. Magnetic field dependence of critical current densities for (c) $H$ $\|$ $c$ and (d) $H$ $\|$ $ab$. The dashed lines show the power-law decaying of $H^{-\alpha}$.}\label{}
\end{figure}
To clarify the fundamental vortex pinning mechanism, we firstly measured magnetic hysteresis loops (MHLs) at several temperatures for both $H$ $\|$ $c$ and $H$ $\|$ $ab$ as shown in Fig. 2(a) and (b), respectively. The loops are almost symmetric, indicating that the bulk pinning is dominating in the crystal. The value of $M$ is monotonically decreasing with increasing $H$ for both directions, namely, the fish-tail effect is absent in FeSe. Such behavior is also reported in the isovalently substituted BaFe$_2$(As$_{1-x}$P$_x$)$_2$.\cite{vanderBeekPRL2010} However, it is quite different from most of other IBSs,\cite{ProzorovPRBBaCoJc,HaberkornPRBCaNa,ShenBingPRB,SalemSuguiPRB,SunAPEX,SunEPL} and previous report,\cite{AbdelPRB} in which the fish-tail effect is observed. We will discuss the fish-tail effect in detail later.

Actually, the absence of fish-tail effect can be witnessed more clearly in the field dependent critical current density, $J_c$, in Fig. 2(c) and (d) obtained by using the extended Bean model:\cite{Beanmodel}
\begin{equation}
\label{eq.1}
J_c=20\frac{\Delta M}{a(1-a/3b)},
\end{equation}
where $\Delta$\emph{M} is \emph{M}$_{down}$ - \emph{M}$_{up}$, \emph{M}$_{up}$ [emu/cm$^3$] and \emph{M}$_{down}$ [emu/cm$^3$] are the magnetization when sweeping fields up and down, respectively, \emph{a} [cm] and \emph{b} [cm] are sample widths (\emph{a} $<$ \emph{b}). In tetragonal two-dimensional systems, there are three kinds of critical current density, \emph{J}$_c^{x,y}$, where \emph{x} and \emph{y} refer to the directions of current and magnetic field, respectively. For \emph{H} $\|$ \emph{c}, irreversible magnetization is determined solely by \emph{J}$_c^{ab,c}$. This means that \emph{J}$_c^{ab,c}$ (= \emph{J}$_c^{H||c}$) can be easily evaluated from the measured magnetization using the extended Bean model. On the other hand, in the case of \emph{H} $\|$ \emph{ab}, both \emph{J}$_c^{ab,ab}$ and \emph{J}$_c^{c,ab}$ contribute to the measured magnetization. Here we simply assume that \emph{J}$_c^{ab,ab}$ is equal to \emph{J}$_c^{c,ab}$, and obtain the weighted average for \emph{H} $\|$ \emph{ab} using eq.(1).\cite{SunAPEX} The self-field \emph{J}$_c$ reaches a value $\sim$ 3 $\times$ 10$^4$ A/cm$^2$ at 2 K for both directions. This value of \emph{J}$_c$ is among the largest values reported so far for FeSe single crystals.\cite{LeiHechangFeSeJc}

When $H$  $\parallel$ $c$, with increasing field, $J_c$ changes little below 1 kOe, followed by a power-law decay $H^{-\alpha}$ at the field of 5 - 10 kOe with $\alpha_1$ $\sim$ 0.5. Such behavior is also observed in most IBSs, which is attributed to strong pinning by sparse nm-sized defects as in the case of YBCO films.\cite{vanderBeekPRB2010,vanderBeekPRB2002} Such a result is consistent with the STM observation, which shows randomly dispersed defects/impurities with effect in nm-sized.\cite{KasaharaPNAS} Above $\sim$ 10 kOe, the decaying rate of $J_c$ increases to $\alpha_2$ $\sim$ 1.2. Such behavior may be caused by the pinning from twin boundaries,\cite{SongPRL,WatashigeSTM} which follow the decrease of $J_c$ $\propto$ $H^{-1}$.\cite{KrusinElbaumPRB} The twin boundaries are usually parallel to $c$-axis, and have little effect when the field is tilted away from $c$-axis. However, field dependent $J_c$ measured for magnetic field at an angular of $\sim$ 20$^\circ$ from $c$-axis shows similar results to that for $H$  $\parallel$ $c$, indicating that the twin boundaries may not be the main reason for $J_c$ $\propto$ $H^{-1}$ dependence. It may be originated from the relatively low density of twin boundaries. Although the twin boundaries are strong pinnings, and will trap vortex for sure, they are easily totally occupied by vortices under a very small field because of their low density. Actually, the density of nm-sized defects/impurities observed by STM is less than one per 2000 Fe atoms.\cite{KasaharaPNAS} In such a case, all the pinning centers (nm-sized defects/impurities) will be occupied by the vortices above a characteristic field of a few tens of kOe. Above that field, the pinning force $F_p$ will keep constant in spite of the increase in $H$. Thus, the value of  $J_c$ will decrease at a rate of $H^{-1}$ since $F_p$ = $\mu_0H$$\cdot$$J_c$. The pinning centers from point defects are mostly randomly distributed in all the three dimensions. Thus, the slope change is also observed when $H$ $\parallel$ $ab$ as shown in Fig. 2(d). However,  the change of $\alpha$ $\sim$ 0.5 to 1 occurs more gradually in the case of $H$  $\parallel$ $ab$. It may suggest the presence of additional pinning mechanisms such as stacking fault parallel to $ab$-plane.

\begin{figure}\center
\includegraphics[width=8.5cm]{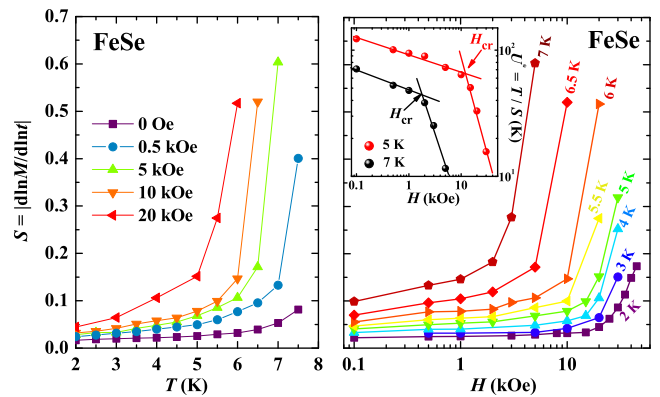}\\
\caption{(Color online) (a) Temperature dependence of magnetic relaxation rate $S$ at different fields. (b) Magnetic field dependence of $S$ at different temperatures. Inset is the double logarithmic plot of the effective pinning energy $U$$^*$ (= $T$/$S$) vs $H$ at 5 and 7 K.}\label{}
\end{figure}
To investigate the vortex dynamics of FeSe, we carefully traced the decaying of magnetization with time $M(t)$ for more than one hour originated from flux creep, where $t$ is the time from the moment when the critical state is prepared. The normalized magnetic relaxation rate $S$ can be obtained from $S$ = $\mid$dln$M$/dln$t$$\mid$. In these measurements, magnetic field was swept more than 5 kOe higher than the target field. Fig. 3(a) shows the temperature dependence of magnetic relaxation rate $S$ at several fields. Obviously, before the steep increase at high temperature, $S$ shows an obvious plateau with a relatively high vortex creep rate (e.g. $S$ $\sim$ 0.02 for zero field). The plateau and large vortex creep rate were also observed in YBa$_2$Cu$_3$O$_{7-\emph{$\delta$}}$, \cite{Yeshurunreview} and other IBSs,\cite{ProzorovPRBBaCoJc,HaberkornPRBBaCo,HaberkornPRBCaNa,ShenBingPRB,TaenPRBBaCoirra,SunEPL,YangHPRB} which can be interpreted by the collective creep theory. \cite{Yeshurunreview}

Fig. 3(b) shows the field dependence of $S$ at temperatures ranging from 2 to 7 K. At small magnetic fields, $S$ slightly increases with applied field. However, the crossover to fast creep occurs at progressively higher fields with lowering temperature. Here, we should point out that the value of $S$ shows monotonic behavior with field, different from YBa$_2$Cu$_3$O$_{7-\emph{$\delta$}}$,\cite{CivalePRB} iron-based "122", \cite{ProzorovPRBBaCoJc,NakajimaPRBirra} and FeTe$_{1-x}$Se$_x$,\cite{SunEPL,TaenPC} in which $S$ at low temperatures shows an upturn with decreasing field to zero. Such an increase in $S$ usually corresponds to the smaller value of $\mu$. As the temperature is lowered, $J_c$ increases leading to the wider distribution of the local field in the sample. When the applied field is not considerably larger than the self-field, local magnetic induction in the region close to the edge of the sample becomes much smaller than the applied field, making this region close to the single vortex regime with smaller $\mu$. The absence of the upturn behavior indicates that the single vortex regime may not exist in the FeSe single crystal or may exist at temperatures lower than the measurement limit 2 K.

\begin{figure}\center
\includegraphics[width=8.5cm]{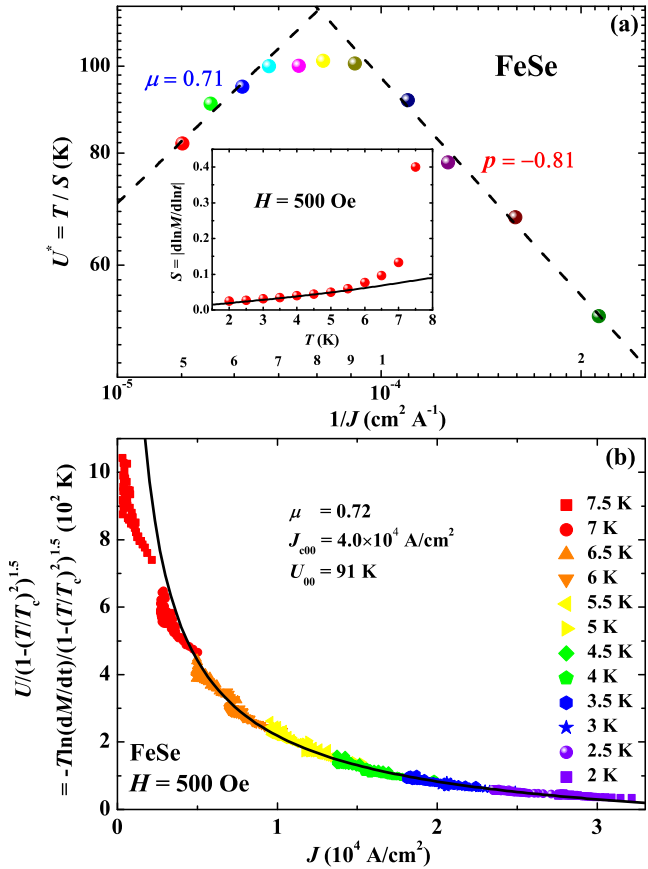}\\
\caption{(Color online) (a) Inverse current density dependence of effective pinning energy \emph{U}$^*$ at 500 Oe in FeSe single crystal. Inset is the temperature dependence of $S$ at 500 Oe. (b) Current density dependence of flux activation energy \emph{U} constructed by the extended Maley's method. The solid line indicates power-law fitting in the large \emph{J} region. }\label{}
\end{figure}

According to the collective creep theory \cite{Yeshurunreview}, magnetic relaxation rate \emph{S} can be described as
\begin{equation}
\label{eq.2}
S=\frac{T}{U_0+\mu Tln(t/t_{eff})},
\end{equation}
where $U_0$ is the temperature-dependent flux activation energy in the absence of flux creep, \emph{t}$_{eff}$ is the effective hopping attempt time, and $\mu$ $>$ 0 is a glassy exponent for elastic creep. The value of $\mu$ contains information about the size of the vortex bundle in the collective creep theory. In a three-dimensional system, it is predicted as $\mu$ = 1/7, (1) 5/2, 7/9 for single-vortex, (intermediate) small-bundle, and large-bundle regimes, respectively.\cite{Blatterreview,FeigelmanPRL}

The flux activation energy \emph{U} as a function of current density \emph{J} can be defined as \cite{FeigelmanPRB}
\begin{equation}
\label{eq.3}
U(J)=\frac{U_0}{\mu}[(J_{c0}/J)^\mu-1].
\end{equation}
Combining this with $U$ = $T$ln($t$/$t_{eff}$) extracted from the Arrhenius relation, we can deduce the so-called interpolation formula
\begin{equation}
\label{eq.4}
J(T,t)=\frac{J_{c0}}{[1+(\mu T/U_0)ln(t/t_{eff})]^{1/\mu}},
\end{equation}
where $J_{c0}$ is the temperature dependent critical current density in the absence of flux creep. From Eqs. (3) and (4), effective pinning energy $U$$^*$ = $T$/$S$ can be derived as
\begin{equation}
\label{eq.5}
U^*=U_0+\mu Tln(t/t_{eff})=U_0(J_{c0}/J)^\mu.
\end{equation}

Thus, the value of $\mu$ can be easily obtained from the slope in the double logarithmic plot of $U$$^*$ vs 1/$J$. Fig. 4(a) shows a typical result at 500 Oe (much larger than the self-field of the FeSe single crystal $\sim$ 100 Oe.). The evaluated value of $\mu$ is $\sim$ 0.71 as expected for collective creep by large bundles similar to that reported in Na-doped CaFe$_2$As$_2$.\cite{HaberkornPRBCaNa} Contrary to the above prediction of $\mu$ $>$ 0, a negative slope with value $\sim$ -0.81 is obtained at small $J$. The negative slope is often denoted as $p$ in the plastic creep theory, which is thought to lead to faster escape of vortices from the superconductors.\cite{AbulafiaPRL} Thus, the FeSe single crystal undergoes a crossover from elastic to plastic creep regimes.

Actually, such crossover from elastic to plastic creep was proposed to be a possible origin of the fish-tail effect. In cuprates and IBSs, the crossover was indeed observed at the same magnetic field as the peak position of the fish-tail.\cite{AbulafiaPRL,ProzorovPRBBaCoJc} However, in our FeSe single crystal, despite the fact that the crossover exists, the fish-tail effect is absent. It indicates that the crossover may have no direct relation with the fish-tail effect, or only the crossover in vortex creep regime may be not enough to cause the fish-tail effect. The absence of fish-tail effect is also observed in the isovalently doped BaFe$_2$(As$_{1-x}$P$_x$)$_2$, which is understood by innately containing only strong pinning sites.\cite{vanderBeekPRL2010} On the contrary, most other IBSs contain weak-pinning sites because of electron/hole doping, such as oxygen vacancies in REFeAsO$_{1-x}$ (RE: rare earth), electron doping Co atom in Ba(Fe$_{0.93}$Co$_{0.07}$)$_2$As$_2$. Thus, the motion of vortices in those crystals is affected by both strong and weak pinnings. These charged quasi-particle-scattering centers are considered as the origin of fish-tail effect.\cite{vanderBeekPRL2010}

FeSe is innately superconducting without further doping, which contains no doping induced charged quasi-particle-scattering centers similar to BaFe$_2$(As$_{1-x}$P$_x$)$_2$. On the other hand, previous STM results show that the main effect of defects in the crystal is in nm-size (with the density less than one defect/impurity per 2000 Fe atoms),\cite{KasaharaPNAS} which will act as strong pinning sites rather than the weak pinning sites induced by atomic sized defects. Thus, the absence of fish-tail effect in FeSe may be attributed to the dominance of strong pinnings. However, we cannot totally exclude the existence of weak pinning centers in FeSe since the collective creep was indeed observed in magnetic relaxation. It may come from slightly non-stoichiometric ratio of Fe to Se. The effect of such a small amount of weak pinning centers from non-stoichiometry is almost negligible compared with that from the strong pinning centers. Thus, the field dependence of critical current density is dominated by the strong pinning. On the other hand, since the contribution to the temperature dependence of $J_c$ from the strong point pinning is expected to be small, the collective pinning  can be observed in $S$ - $T$ curves. Our results indicate that the emergence of fish-tail effect needs comparable strong and weak pinnings. Introduction of weak pinning centers into FeSe by particle irradiations such as using  electrons may induce the fish-tail effect.

Here, we should point out that weak fish-tail effect was observed in a related material FeTe$_{1-x}$Se$_x$,\cite{SunAPEX,SunEPL} which is also an isovalently doped system. This exception may come from the sample quality, in which the interstitial Fe (excess Fe) in FeTe$_{1-x}$Se$_x$ may work as the charged quasi-particle-scattering centers. Actually the interstitial Fe has been proved to be in the valence state near Fe$^+$, and the amount can be as large as 14\%, which can contribute as weak pinning centers.\cite{ZhangLijunPRB,SunSciRep}

In the following, we analyze the $U$ - $J$ relation by the extended Maley's method, which considers the temperature dependence of $U$ into the original Maley's method.\cite{MiuMayle} This method allows to scale $U$ in a wide range of $J$. The temperature dependent $U_0$ is assumed as $U_0(T) = U_{00}[1-(T/T_c)^2]^n$. Here, the exponent $n$ is set to 3/2 as in the case of  YBa$_2$Cu$_3$O$_{7-\emph{$\delta$}}$ \cite{ThompsonPRB,MiuMayle}, Co-doped BaFe$_2$As$_2$,\cite{TaenPRBBaCoirra} and FeTe$_{1-x}$Se$_x$.\cite{SunEPL} It is obvious that all the curves can be well scaled together without introducing any more factors as shown in Fig. 4(b). The solid line indicates the power-law fitting by Eq. (3) to the large $J$ region above $\sim$ 1 $\times$ 10$^4$ A/cm$^2$ where the slope in Fig. 4(a) is positive. Deviation of the data from the fitting line in the small $J$ region is reasonable since vortex creep is plastic there. The fitting gives independent evaluation of $\mu$ = 0.72, activation energy $U_{00}$ = 91 K, and $J_{c00}$ = 4.0 $\times$ 10$^4$ A/cm$^2$. The value of glassy exponent obtained from the extended Maley's method is very close to that evacuated in Fig. 4(a). With the value of $U_{00}$, temperature dependence of $S$ is fitted by Eq. (2) with a single free parameter of $\mu$ln($t$/$t_{eff}$) = 10 shown as the solid line in the inset of Fig. 4(a).

Finally, with the above results, we constructed a vortex phase diagram for FeSe single crystal as shown in Fig. 5. $H_{c2}$ represents the putative upper critical field obtained from the midpoint of the resistivity transition at $T_c$ under fields up to 90 kOe. Here we should point out that the determination of $H_{c2}$ may be ambiguous because of the large superconducting fluctuations.\cite{KasaharaPNAS} $H_{irr}$ is the irreversibility field obtained by extrapolating $J_c$ to zero in $J_c^{1/2}$ vs $H$ curves. $H_{cr}$ is the crossover field from elastic to plastic creep, which is obtained from double logarithmic plot of the curves of $U$$^*$ (= $T$/$S$) vs $H$ at different temperatures as shown in the inset of Fig. 3(b). Obviously, at small magnetic fields, vortices in FeSe elastically (collectively) creep in the form of vortex bundle at a relatively low speed. With increasing field over $H_{cr}$, the vortex creep suddenly becomes faster, and enters the plastic creep regime. Further increasing field over $H_{irr}$, the motion of vortices changes from  creeping in solid state to an unpinned liquid state.

\begin{figure}\center
\includegraphics[width=8.5cm]{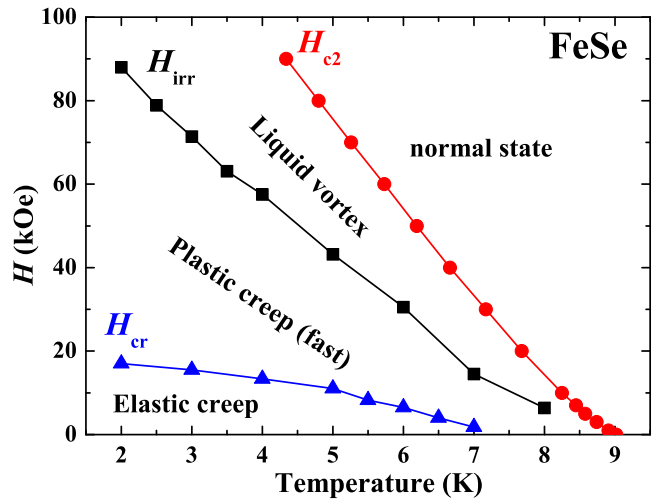}\\
\caption{(Color online) The vortex phase diagram for FeSe single crystal with $H$ $\|$ $c$.}\label{}
\end{figure}

\section{conclusions}
In summary, we have studied the vortex dynamics in high-quality FeSe single crystal by performing magnetization measurements of the critical current density and magnetic relaxation rate. The higher $T_c$, small $\Delta$$T_c$, and large RRR manifest the very high-quality of our crystal. Critical current density, $J_c$, is proved almost isotropic, and reaches a value $\sim$ 3 $\times$ 10$^4$ A/cm$^2$ at 2 K (self-field) for both $H$ $\|$ $c$ and $ab$. Magnetic relaxation rate $S$ shows a temperature insensitive plateau behavior in the intermediate temperature region, which is interpreted in the framework of the collective creep theory. A crossover from elastic to plastic creep is observed, despite the fact that the fish-tail effect is absent for both $H$ $\|$ $c$ and $ab$. Based on this, origin of the fish-tail effect is also discussed. Combining the results of $J_c$ and $S$, vortex motion in FeSe single crystal is proved to be dominated by sparse strong point-like pinning from nm-sized defects. The weak collective pinning is also observed and proved in the form of large bundles. Besides, the vortex phase diagram of FeSe was also constructed.
\acknowledgements
We thank T. Taen for helpful discussion. Y.S. gratefully appreciates the support from Japan Society for the Promotion of Science. This work is partly supported by JSPS/MEXT KAKENHI (Grants-in-Aid for Scientific Research), Japan.

%\bibliography{references}% Produces the bibliography via BibTeX.
\bibliography{references}

\end{document}